\documentclass[12pt]{iopart}
\usepackage{iopams}
\usepackage{graphicx}
\usepackage{color}

\newcommand{\rydn}{\ensuremath{n}}
\begin{document}

\title[Rabi oscillations between ground and Rydberg states]{Rabi oscillations between ground and Rydberg states and van der Waals blockade in a mesoscopic frozen Rydberg gas}

\author{M Reetz-Lamour, J Deiglmayr, T Amthor and M~Weidem\"{u}ller}

\address{Physikalisches Institut, Universit\"{a}t Freiburg, Hermann-Herder-Str. 3, D-79104 Freiburg, Germany}
\vspace*{5pt}\address{Online at: {\tt
http://quantendynamik.physik.uni-freiburg.de}}

\ead{\mailto{m.weidemueller@physik.uni-freiburg.de}}

\begin{abstract}
We present a detailed analysis of our recent observation of
synchronous Rabi oscillations between the electronic ground state
and Rydberg states in a mesoscopic ensemble containing roughly 100
ultracold atoms [M. Reetz-Lamour \textit{et al.}, submitted,
arXiv:0711.4321]. The mesoscopic cloud is selected out of a sample
of laser-cooled Rb atoms by optical pumping. The atoms are coupled
to a Rydberg state with principal quantum number around 30 by a
two-photon scheme employing flat-top laser beams. The influence of
residual spatial intensity fluctuations as well as sources of
decoherence such as redistribution to other states, radiative
lifetime, and laser bandwidth are analysed. The results open up
new possibilities for the investigation of coherent many-body
phenomena in dipolar Rydberg gases. As an example we demonstrate
the van der Waals blockade, a variant of the dipole blockade, for
a mesoscopic atom sample.
\end{abstract}


\section{Introduction: On aims and obstacles}

Since the early days of atomic physics, atoms in highly excited
electronic states (``Rydberg atoms'') have been extensively
studied due to their exaggerated properties marking the borderline
between classical and quantum physics~\cite{gallagher94}.
Recently, Rydberg atoms have attracted interest as a possible
implementation of quantum information processing based on neutral
atoms~\cite{jaksch00,lukin01,miroshnychenko06,cozzini06}.
Representing an alternative to trapped ions or nuclear spins in
molecular complexes~\cite{bouwmeester00,nielsen00}, neutral atoms
offer extraordinarily weak coupling to dissipative processes of
the environment thus promising long coherence
times~\cite{treutlein04}. Several schemes have been proposed to
entangle neutral atoms and realise quantum
gates~\cite{turchette95,jaksch99,lukin01}. Ensembles of ultracold
Rydberg atoms appear to be particularly promising as they exhibit
controllable electric dipole interactions over distances of many
micrometres. The coherent character of these dipolar
Rydberg-Rydberg interactions has been explored  in ultracold gases
over the last
years~\cite{anderson98,mourachko98,anderson02,mudrich05,westermann06}
paving the way for the implementation of fast two-qubit gates.

One prominent example for the strong Rydberg-Rydberg interaction
is the so-called van der Waals blockade. The van der Waals
blockade is the inhibition of multiple excitations in a mesoscopic
ensemble due the interaction-induced energy shifts which separates
the single excitation from multiple excitations energetically. It
is a favourable tool in quantum information
protocols~\cite{lukin01} and has resulted in a number of
experimental investigations. It was first observed as a
suppression of excitation in macroscopic clouds for strong van der
Waals interaction at high principal quantum
numbers~\cite{tong04,singer04}, later for both resonant and
permanent dipoles~\cite{vogt06,vogt07}. It is accompanied by a
change in counting statistics~\cite{cubel05a} and a modified
many-body Rabi frequency which has not been observed directly so
far but evidence of which has recently been
found~\cite{heidemann07}.

Despite this success, the realisation of single qubit operations
with Rydberg atoms has proven to be particularly demanding. While
stimulated adiabatic passage has been used to achieve high Rydberg
excitation probabilities~\cite{cubel05,deiglmayr06}, Rabi
oscillations of atoms starting from the ground state are hard to
observe due to the small spatial overlap between the electronic
wavefunctions of ground and highly excited states resulting in
very small dipole transition matrix elements. In order to achieve
sufficiently high Rabi frequencies despite the small dipole matrix
elements between the electronic ground state and Rydberg states,
one therefore needs high laser intensities provided by tightly
focussed beams. The Gaussian beam profile of a focussed laser
beam, however, exhibits a wide range of intensities and would
therefore impede the observation of synchronous Rabi floppings in
a mesoscopic sample~\cite{deiglmayr06}. One may overcome this by
preparing the atomic cloud in very small dipole traps, rendering
the atomic cloud smaller than the excitation laser and thus only
using the central part of a focussed Gaussian beam. This technique
has recently been used for clouds between 1 and 10
atoms~\cite{johnson07}.

We have recently demonstrated Rabi oscillations between ground and
Rydberg states in a frozen mesoscopic sample of roughly 100
ultracold atoms by applying beam shaping to ensure a constant
intensity distribution over the excitation volume~\cite{reetz07}.
We restrict the excitation volume to the area where the shaped
beam has a sufficiently constant intensity by spatially selective
optical pumping. In this way we filter a signal of a subensemble
of only 100 atoms out of a cloud of ~10 million atoms. Here we
discuss the experimental details and thoroughly investigate
important sources of decoherence. In addition we demonstrate the
van der Waals blockade for this mesoscopic system by observing a
density-dependent suppression of excitation for strong
van-der-Waals interaction at large principle quantum numbers
\rydn.

\section{Experimental realisation}

\begin{figure}
 \begin{indented}
  \item[]
  \includegraphics[width=.75\textwidth]{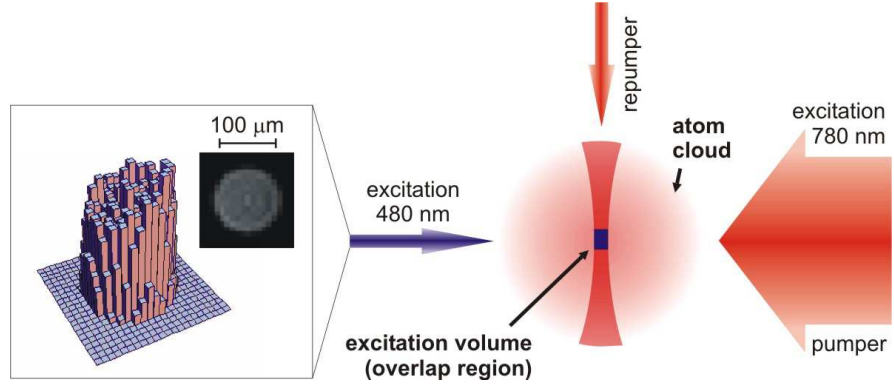}
 \end{indented}
\caption{Preparation of a mesoscopic subensemble through spatially
selective optical pumping and Rydberg excitation with a spatially
homogeneous laser intensity. A small tube of atoms within a cloud
of $10^7$ ultracold atoms is marked by optical pumping with the
pumper and repumper laser beam. A mesoscopic subensemble
containing about 100 atoms within this tube is transferred to a
Rydberg state by near-resonant two-photon excitation with two
counter-propagating laser beams at 780\,nm and 480\,nm,
respectively. The excitation laser at 480\,nm has a flat-top
intensity profile shown in the inset, while the beam at 780\,nm
superimposed on the pumper beam has a Gaussian profile as all
other beams.}
 \label{fig:setup}
\end{figure}

\begin{figure}
 \begin{indented}
  \item[]\includegraphics[width=.5\textwidth]{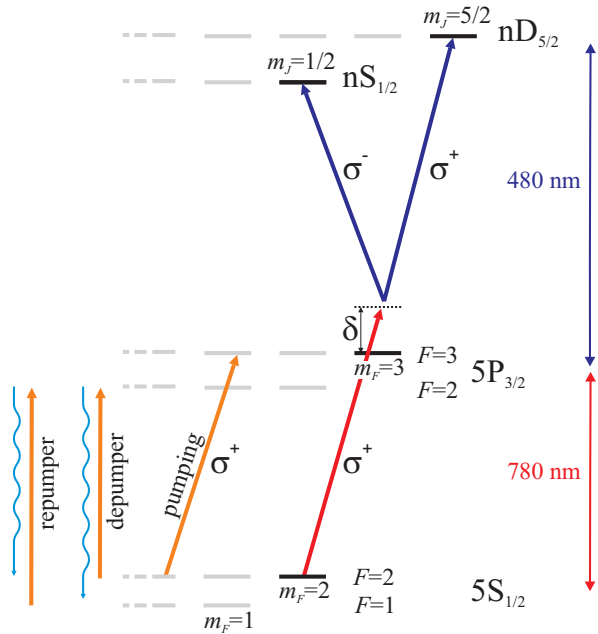}
 \end{indented}
\caption{Level scheme for optical pumping and near-resonant
two-photon excitation of $^{87}$Rb Rydberg states. Addressing of a
mesoscopic subset of atoms involves depumping, spatially selective
repumping, and optical pumping in order to single out the 3-level
subsystem $\left|5\mathrm{S}_{1/2},F=2,m_F=2\right>
\;\rightarrow\; \left|5\mathrm{P}_{3/2},F=3,m_F=3\right>
\;\rightarrow\; \left|\rydn\ell_{J=\ell+1/2},m_J=J\right>$, with
$\ell$=0(S) or 2(D). Decoherence effects from the short lifetime
of the intermediate state are minimised by the detuning $\delta$
from the intermediate resonance thereby reducing the system to an
effective two-level system.}
 \label{fig:levelscheme}
\end{figure}

The experiment is performed with a cloud of ultracold atoms in a
vapour-cell magneto-optical trap (MOT) with about $10^7$ $^{87}$Rb
atoms at densities of $10^{10}\,\mathrm{cm}^{-3}$ and a
temperature below 100\,$\mu$K. The excitation is achieved with two
counter-propagating laser beams at 780\,nm and 480\,nm via
$$
 \centering
 5\mathrm{S}_{1/2}
 \stackrel{780\,\mathrm{nm}}{\longrightarrow}
 5\mathrm{P}_{3/2}
 \stackrel{480\,\mathrm{nm}}{\longrightarrow}
 \rydn \mathrm{S_{1/2}}\,/\,\rydn \mathrm{D}_{5/2}
$$
(see Figures~\ref{fig:setup} and~\ref{fig:levelscheme}). By
detuning from the intermediate state $5\mathrm{P}_{3/2}$, this
state experiences negligible population which assures sufficiently
long coherence times ~\cite{deiglmayr06}. The excitation laser at
780\,nm is collimated to a waist of 1.1\,mm ensuring a constant
Rabi frequency over the excitation volume. The Rabi frequency is
determined from measured Autler-Townes
splittings~\cite{autler55,teo03,deiglmayr06} with typical values
of $\omega/2\pi = 55\,$MHz.

The 480\,nm laser alignment is discussed in
Sections~\ref{sec:shaping} and \ref{sec:align}. Its frequency is
locked to a temperature stabilised resonator built from
\textsc{Zerodur}$^{\footnotesize \circledR}$ which has a residual
drift of $\sim$2\,MHz per hour. We compensate for this by
repeating identical spectral measurements, determining the line
centres and interpolating the resonator drift.

All lasers are switched by accousto-optical modulators (switching
times $\sim$50\,ns) and guided to the experiment by single-mode
polarisation-maintaining optical fibres. This ensures an unaltered
alignment independent of changes on the input side, \textit{e.g.}
when changing the laser wavelength for excitation of different
Rydberg states or for different detunings from the intermediate
state.

Typically 30\,$\mu$s after excitation all Rydberg atoms are
field-ionised (field pulse height 770\,V/cm) and detected with a
microchannel plate detector whose signal is recorded with a boxcar
integrator. The delay of 30\,$\mu$s between excitation and
detection suppresses a background of ``hot'' (300\,K) atoms while
it has no influence on the measured excitation probability of the
cold subensemble (see \Sref{sec:bg}). We have verified that no
spurious ions are present which is as expected for the short
timescales and small atom numbers considered throughout this
article. After detection the MOT is turned back on and the whole
cycle is repeated every 70\,ms.

\subsection{Beam shaping}
\label{sec:shaping}

\begin{figure}
 \begin{indented}
  \item[] \includegraphics[width=.84\textwidth]{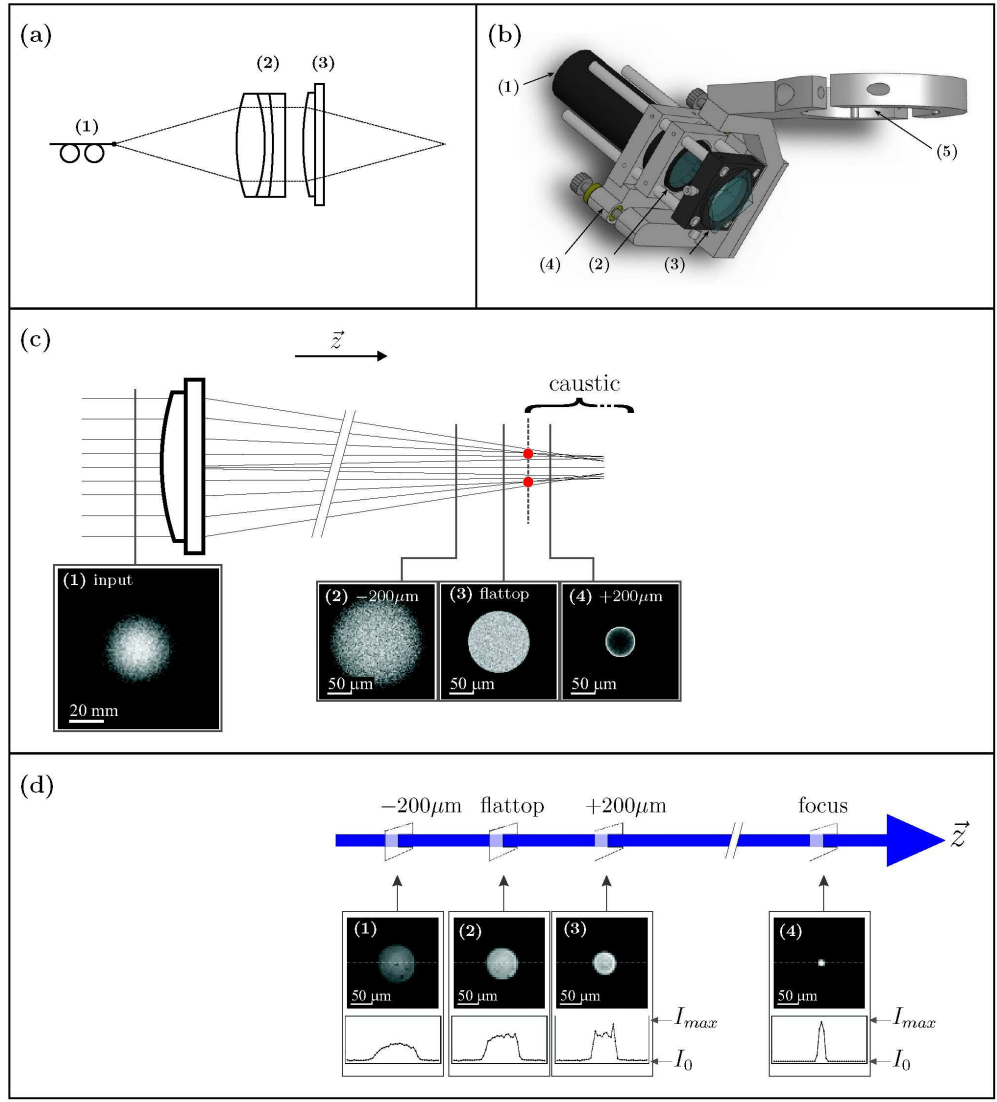}
 \end{indented}
\caption{Realisation of a flattop beam profile for the 480\,nm
excitation laser. \textbf{(a)} Schematic: The output of a
single-mode polarisation-maintaining fibre (1) is collimated by an
achromat to a $1/\mathrm{e}^2$-intensity radius of 6.9\,mm. The
beam is focussed and shaped by a diffractive optical element which
is cemented onto a standard plano-convex lens (3). \textbf{(b)}
Supporting cage with (1) fibre connector, (2) achromat, and (3)
diffractive element which is mounted on a 2-axes translation stage
shown in black. The cage is fixed on an adapted 3-axes mirror
mount (4) which is fixed to the vacuum chamber (5). \textbf{(c)}
Beam shaping principle with ray optics. The collimated input beam
(1) is focussed and shaped into a flattop beam profile at one
specific distance (3). Before this region the shape is closer to a
Gaussian beam (2), behind this point crossing rays (caustics)
cause an increased intensity in the beam edges. \textbf{(d)}
Measured beam profiles: (2) best resemblance of a flattop, (1) and
(3) profiles at a distance of 200\,$\mu$m from the flattop. The
actual focus (4) is 640\,$\mu$m further along the beam and is used
for alignment (see Section~\ref{sec:align}). Figs. c.(4) and d.(3)
show a difference between the designed and realized intensity
distribution. The discrepancy is a consequence of imperfections in
the actual setup, \textit{e.g.} slight misalignment, inaccurate
diameter of input beam, imperfect diffractive element.}
 \label{fig:DOE}
\end{figure}

The 480\,nm beam is focused and shaped to a constant intensity
distribution over the excitation volume: The output of a
single-mode polarisation-maintaining fibre is collimated by an
achromat (focal length 80\,mm, 1 inch diameter) to a beam size
($1/\mathrm{e}^2$ intensity radius) of 6.9\,mm and shaped by a
diffractive optical element (DOE) which is cemented onto a
plano-convex lens with a focal length of 100\,mm (see
Fig.~\ref{fig:DOE}.(a)). The DOE effectively distorts the
plano-convex lens to give a flattop profile at one specific
distance close to the actual focus of the beam. All elements from
the fibre coupler to the DOE are mounted in a  rigid cage that can
be aligned as a whole with respect to the MOT (see
Fig.~\ref{fig:DOE}.(b)).

Fig.~\ref{fig:DOE}.(c) shows the calculated behaviour of the DOE.
The flattop is realised only at one specific distance from the
DOE. Before this point the beam profile is closer to a Gaussian
beam, behind this point some of the beam rays cross (so-called
caustics) which leads to an increased intensity at the beam edges.
Even further along the beam is the actual focus.

We have aligned the elements within the cage by adjusting the
distance between fibre output and achromat as well as the position
of the DOE with respect to the centre of the collimated beam, and
characterised the beam shape with a USB webcam (Philips ToUCam Pro
PCVC740K). The CCD chip has been replaced by a black-and-white
version (ICX098BL) with the protective glass cover removed and we
have modified the firmware (WcRmac, v1.0.79, 2005,
\verb"http://www.astrosurf.com/astrobond/ebrawe.htm"). In this way
we obtain raw, uncompressed, and unprocessed images at the full
spatial resolution of the CCD chip (5.6\,$\mu$m). The actual
performance of the DOE as well as the closest approximation to a
flattop is depicted in Fig.~\ref{fig:DOE}.(d). It is realised at a
distance of 640\,$\mu$m in front of the focal point.

\subsection{Beam alignment: Selection of a mesoscopic subensemble}
\label{sec:align}

\begin{figure}
 \begin{indented}
  \item[]\includegraphics[width=.65\textwidth]{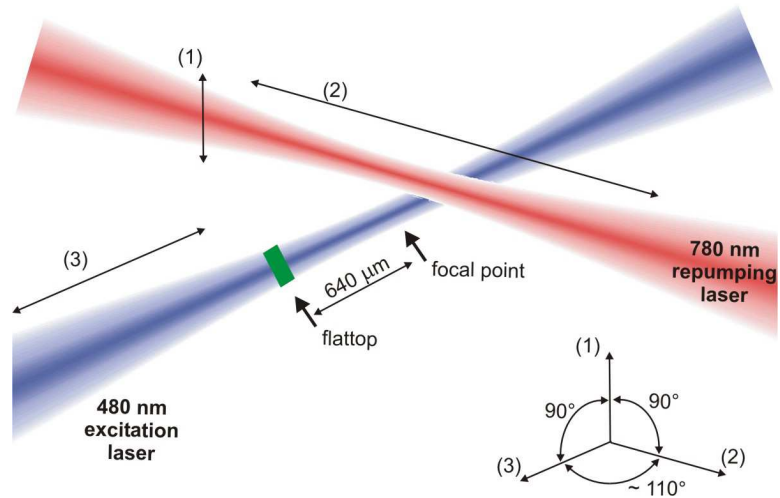}
 \end{indented}
\caption{The excitation volume is given by the overlap region of a
tightly focussed repumping beam (red) and the shaped 480\,nm beam
(blue). The lasers are aligned by first overlapping both foci
(minimising the excitation volume) and then moving the flattop
region onto the red beam by moving the blue focus by 640\,$\mu$m
along the beam direction (see Fig.~\ref{fig:DOE}(d)). Details see
text.}
 \label{fig:align}
\end{figure}

The previous characterisation of the shaped beam allows us to
confine the excitation volume to the flattop region with a
dedicated optical pumping scheme: The magnetic field of the MOT is
turned off 3\,ms before the trapping lasers are switched off and
all atoms are pumped to the dark
$\left|5\mathrm{S}_{1/2},F=1\right>$ ground state within
300\,$\mu$s by a depumping laser (see Fig.~\ref{fig:levelscheme}).
All atoms in a small cylindrical tube perpendicular to the
excitation beams are transferred to the upper hyperfine component
$\left|5\mathrm{S}_{1/2},F=2 \right>$ of the ground state with a
1\,$\mu$s pulse of a tightly focussed repumping beam (waist
10\,$\mu$m, power 10\,nW). All atoms in this tube are optically
pumped to the stretched ``launch'' state
$\left|5\mathrm{S}_{1/2},F=2,m_\mathrm{F}=2\right>$ by a 1$\mu$s
pulse of a $\sigma^+$-polarised pumper beam superimposed with the
780\,nm excitation laser. A small guiding field of 35\,mG along
the axis of the excitation lasers is turned on at all times. A
sketch of the laser beam geometry is shown in \Fref{fig:setup}.
Finally the atoms in the ``launch'' state are excited into Rydberg
states. The excitation volume is therefore given by the overlap
between the 480\,nm laser and the repumping laser. This allows us
to align the lasers by first overlapping both foci (minimising the
excitation volume) and then moving the repumping beam onto the
flattop in the following way (see Fig.~\ref{fig:align}):

The red beam is produced from the output of a single-mode fibre
with a mounting that allows alignment along the directions (1) and
(2). After a rough alignment the focus of the repumping beam is
moved slightly above the blue beam by a small separation in
direction (1). Moving the focus along direction (2) we only get an
overlap with the 480\,nm beam (and thus a Rydberg signal), when
the divergence of the repumping beam is larger than the beam
separation. In this way we can probe the Rayleigh-range of the red
laser and thus find its focal point. After this we can measure the
local width of the blue beam by moving the red laser along
direction (1). As the mounting of the 480\,nm beam allows movement
of its focus along direction (3), we measure the 480\,nm beam
diameter in this way at several positions along direction (3) and
thus overlap the 480\,nm beam focus with the red beam. Finally we
move the flattop region onto the red beam by moving the blue focus
by 640\,$\mu$m along the beam direction (see
Fig.~\ref{fig:DOE}(d)). We estimate an alignment accuracy of
50\,$\mu$m in direction (3) and an accuracy on the order of
10\,$\mu$m in direction (1). The excitation volume is given by the
beam diameters resulting in a volume of about $10\,\mu\mathrm{m}
\times 10\,\mu\mathrm{m} \times 100\,\mu\mathrm{m}$ which (at our
density of $10^{10}\,\mathrm{cm}^{-3}$) corresponds to 100 atoms
in the excitation volume.

\subsection{Background suppression}
\label{sec:bg}

\begin{figure}
 \begin{indented}
  \item[]\includegraphics[width=.65\textwidth]{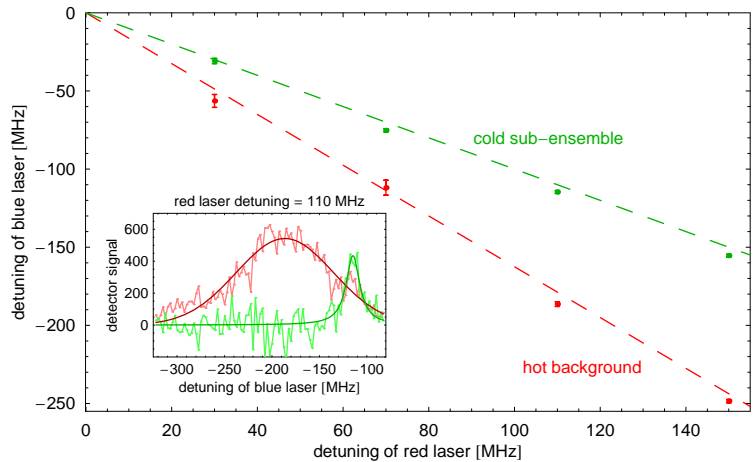}
 \end{indented}
\caption{(Inset) Spectral dependence of both the background signal
determined with a blocked repumping beam (red, with Gaussian fit)
and the signal from the mesoscopic sample (green, with Lorentzian
fit, background already subtracted) for a detuning of
$\delta=2\pi\times110\,$MHz from the intermediate level (see
Fig.~\ref{fig:levelscheme}). For the detunings used in this
article the background signal is usually between one tenth and one
half the size of the signal from the mesoscopic cold ensemble.
(Main graph) Peak position for the background signal (red dots)
for different values of $\delta$ together with the peak position
of the mesoscopic signal (green dots). The dashed green line shows
the expected dependence for a coherent two-photon excitation, the
dashed red line corresponds to the expectation for a hot gas as
given by Eq.~(\ref{eq:bg}). Where error bars are not visible they
are smaller than the dot size.}
 \label{fig:background}
\end{figure}

As our excitation volume is defined by the overlap of the
repumping and the 480\,nm beam (see Fig.~\ref{fig:setup}), we can
determine a background signal by blocking the repumping beam. The
inset of Fig.~\ref{fig:background} shows the spectral dependence
of both the background signal (red) and the signal from the
flattop region (green) on the detuning $\Delta$ of the 480\,nm
laser. The detuning of the 780\,nm laser from the intermediate
5P$_{3/2}, F=3$ state is fixed at $\delta=+2\pi\times110\,$MHz.
The signal from the flattop region peaks at $\Delta \cong
-\delta$, \textit{i.e.} the two-photon resonance condition is
fulfilled, which is the signature of coherent
excitation~\cite{deiglmayr06}. By contrast, the background signal
peaks at a different detuning. Fig.~\ref{fig:background} shows the
dependence of the peak positions for different values of $\delta$.
The spectral dependence shows that the background signal comes
from the hot background gas from which the cold cloud is loaded:
Due to their thermal velocity hot atoms experience the excitation
lasers at their Doppler-shifted frequency. For atoms with a
projected velocity $v$ in the direction of the 780\,nm laser, this
laser is resonant with the transition
5S$_{1/2}\rightarrow$5P$_{3/2}$ if the Doppler shift $-2\pi\times
v\lambda$ is equal to $\delta$, with $\lambda=780\,$nm being the
laser wavelength. As the two excitation lasers are
counter-propagating these atoms are then resonant with the 480\,nm
laser if it is detuned from the resonance by
\begin{equation}
 \label{eq:bg}
 \Delta=2\pi\times\frac{v}{480\,\textrm{nm}}=-\frac{780\,\textrm{nm}}{480\,\textrm{nm}}\,\delta\,.
\end{equation}
The dashed red line in Fig.~\ref{fig:background} shows this
dependence in good agreement with the experimental values for the
background spectra.

In this way the thermal velocity can be used to separate the
signal from the hot atoms spectroscopically from the signal of the
cooled atoms in the flattop excitation volume. We reduce the
background even further by delaying the detection: After each
excitation we wait for 30\,$\mu$s before detection. In this time
most of the hot atoms have already left the detection region which
allows for a significant background suppression. Longer delays are
not possible as the lifetime of the states discussed here are of
the same order (\textit{e.g.} 26\,$\mu$s for 35D).

All figures shown in this article have been acquired by
subtracting the background of hot atoms, which is determined by
repeating the corresponding measurement with a blocked repumping
beam.

\subsection{Observation of Rabi oscillations}

\begin{figure}
 \begin{indented}
  \item[]\includegraphics[width=.65\textwidth]{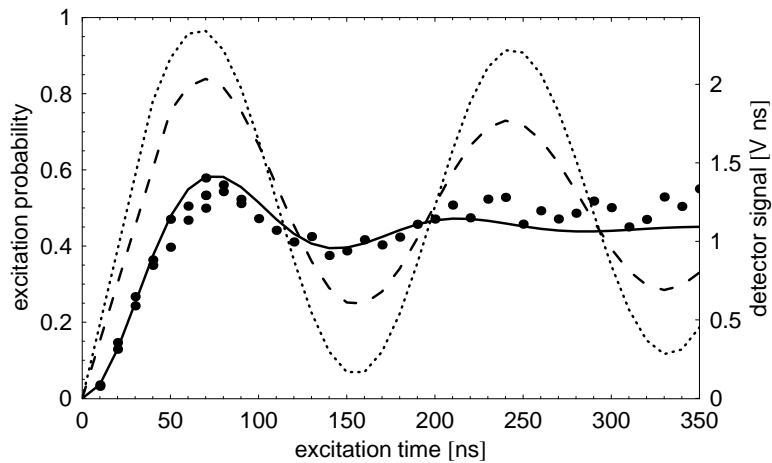}
 \end{indented}
\caption{Rabi oscillations of atoms excited to the 31D$_{5/2}$
state at a detuning from the intermediate state of
$\delta\,/\,2\pi=140\,$MHz and a Rabi frequency for the lower
transition of $\omega\,/\,2\pi = 55\,$MHz. Each dot corresponds to
an average of measurements over 28 experimental repetition cycles.
The dotted line is the simulated excitation probability for a
perfect flattop, \textit{i.e.} a single Rabi frequency. In this
case the damping comes from a small but finite admixture of the
short-lived intermediate 5P$_{3/2}$ state (see
Sec.~\ref{sec:dephasing}). The dashed line shows a simulation that
additionally averages over the measured intensity distribution.
Finally the solid line also takes an effective laser linewidth of
2.4\,MHz into account. The MCP signal is given in V\,ns, and the
axis is scaled to overlap with the simulated curve. The scaling
factor constitutes the detector efficiency measured in mV\,ns per
atom. The only free parameter for the simulations is the Rabi
frequency for the upper transition which is compared to ab-initio
calculations in Sec.~\ref{sec:abinitio}.}
 \label{fig:expRabi}
\end{figure}

Our excitation scheme allows to excite either \rydn S or \rydn D
states (\rydn~denotes the principal quantum number) depending on
the helicity of the polarisation: We excite atoms to \rydn S
states via
$$
 \centering
 \left|5\mathrm{S}_{1/2},F=2,m_F=2\right>
 \stackrel{\sigma^+}{\longrightarrow}
 \left|5\mathrm{P}_{3/2},F=3,m_F=3\right>
 \stackrel{\sigma^-}{\longrightarrow}
 \left|\rydn \mathrm{S}_{1/2},m_J=1/2\right>
$$
and \rydn D states via
$$
 \centering
 \left|5\mathrm{S}_{1/2},F=2,m_F=2\right>
 \stackrel{\sigma^+}{\longrightarrow}
 \left|5\mathrm{P}_{3/2},F=3,m_F=3\right>
 \stackrel{\sigma^+}{\longrightarrow}
 \left|\rydn \mathrm{D}_{5/2},m_J=5/2\right>
$$
as depicted in Fig.~\ref{fig:levelscheme}.\footnote{Note that
$\left|5\mathrm{P}_{3/2},F=3,m_F=3\right> =
\left|5\mathrm{P}_{3/2},I=3/2,m_I=3/2,m_J=3/2\right>.$}

In Fig.~\ref{fig:expRabi} the measured fraction of excited Rydberg
atoms in the 31D$_{5/2}$ state as a function of excitation time is
shown. The excitation time is given by the pulse length of the
480\,nm laser, as the 780\,nm light is switched on (off) 100\,ns
before (after) the 480\,nm laser to ensure constant intensity
during the on-time of the latter. Fig.~\ref{fig:expRabi} shows a
measured Rabi oscillation. It is damped out quickly. For times
larger than 200 ns the number of Rydberg atoms reaches a steady
state which indicates the balance between excitation and
down-stimulation. The same temporal dependence is observed for
\rydn S and \rydn D states with $\rydn \lesssim 40$. For these
small \rydn~it does not depend on the atom density. This indicates
that the observed Rabi oscillations are not affected by
Rydberg--Rydberg interactions and that all atoms simultaneously
perform identical Rabi oscillations. As seen in
Fig.~\ref{fig:expRabi} the oscillations are strongly damped, which
is a result of the remaining admixture of the intermediate state,
the residual intensity distribution of the flat-top beam profile,
and our finite laser bandwidth. We will discuss these sources for
decoherence in Section~\ref{sec:dephasing}.

\section{Systematics}

\subsection{Dependence on the detuning from the intermediate state}

\begin{figure}
 \begin{indented}
  \item[]\includegraphics[width=.65\textwidth]{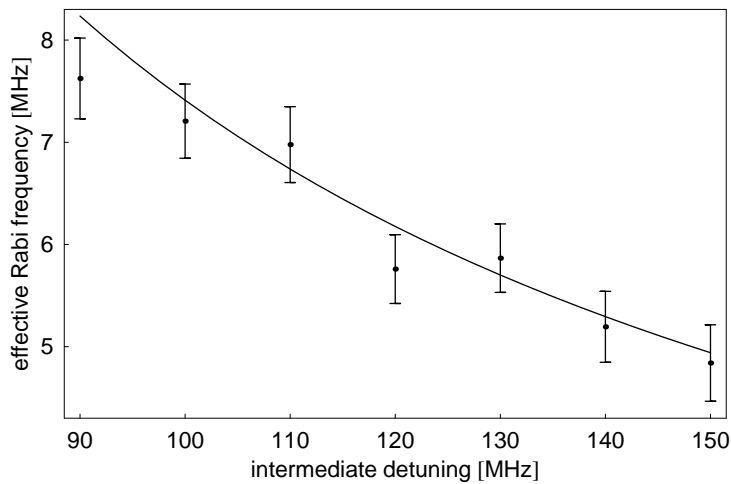}
 \end{indented}
\caption{Measured effective Rabi frequencies
$\Omega_{\mathrm{eff}}\,/\,2\pi$ for different detunings
$\delta\,/\,2\pi$ from the intermediate state. The solid line
shows the theoretical prediction for a blue Rabi frequency
$\Omega\,/\,2\pi = 27\,$MHz and a red Rabi frequency
$\omega\,/\,2\pi = 55\,$MHz as deduced from an Autler-Townes
splitting.}
 \label{fig:detuning}
\end{figure}

If the detuning from the intermediate level is sufficiently large,
one can reduce the actual three-level system to a two-level system
with an effective Rabi frequency of $\Omega_{\mathrm{eff}} =
\omega\,\Omega\,/\,2\delta$ where $\omega$ ($\Omega$) denotes the
Rabi frequency of the lower (upper) transition and $\delta$
denotes the detuning from the intermediate level. We have verified
this dependence by measuring the effective Rabi frequencies for
different detunings from the intermediate level. The effective
Rabi frequencies are determined by a fit of a damped oscillation,
$\frac{1}{2}\left[1-\exp(-t/\tau_{\mathrm{damp}})\,\cos(\Omega_{\mathrm{eff}}\,t)\right]\,,\;\tau_{\mathrm{damp}}=60\,\mathrm{ns}$,
to measured Rabi oscillations. The resulting fit values together
with the fit uncertainties are depicted in
Fig.~\ref{fig:detuning}. They are in very good agreement with the
expectation for an effective two-level system.

\subsection{Comparison to ab-initio transition matrix elements}
\label{sec:abinitio}

\begin{table}
\caption{Comparison between measured Rabi frequencies and
calculations based on ab-initio transition matrix elements.}
\label{tbl:TheoVsExp}
\begin{indented}
\item[]
\begin{tabular}{ccc}
\br
final state & $\Omega_{\mathrm{exp}}\,/\,2\pi$ & $\Omega_{\mathrm{theo}}\,/\,2\pi$\\
\mr
30D$_{5/2}$ & 32.8$\,\pm\,$1.7\,MHz & 37.6\,MHz\\
31S$_{1/2}$ & 18.9$\,\pm\,$0.6\,MHz & 19.3\,MHz\\
\br
\end{tabular}
\end{indented}
\end{table}

The experimental values of the Rabi frequencies are compared to
ab-initio calculations yielding a concise test of predicted
transition matrix elements. The Rabi frequency scales with the
blue laser intensity $I$ as $\Omega \propto \mu\,\sqrt{I}$ where
$\mu$ is the matrix element between the states coupled by the
laser light, \textit{i.e.} the $5\mathrm{P}_{3/2}$ and the Rydberg
state.

To determine $\mu$ we split the Rydberg electron wavefunctions
$\psi_{\rydn\,\ell\,j\,m_j}$ into a radial and a spherical part
with the ansatz
\begin{equation}
\label{eq:totalWavefunction}
 \psi_{\rydn\,\ell\,j\,m_j}
 = \sum\limits_{m_s}
 \frac{1}{r}U_{\rydn\,\ell\,j}(r)\,
 Y_{\ell\,m_j-m_s}(\theta,\varphi)\,
 C^{j,m_j}_{\ell,m_j-m_s,\frac{1}{2},m_s}\,
 \Theta^{1/2}_{m_s},
\end{equation}
where $Y_{\ell\,m_\ell}$ denote the well-known spherical
harmonics, $C^{j,m_j}_{\ell,m_j-m_s,\frac{1}{2},m_s}$ is the
Clebsch-Gordan coefficient and $\Theta^{1/2}_{m_s}$ is the spin
wavefunction. The radial part is obtained by numerically
integrating the Schr\"odinger equation with the Numerov algorithm
following Ref.~\cite{zimmerman79} at the energy $E_{\rydn \ell j}$
(calculated with the according quantum defect). To obtain
reasonable results for low-lying levels we use a model potential
that is fitted to one-electron energies~\cite{marinescu94}. The
knowledge of the radial wavefunctions allows us to calculate the
radial part $\mu_{\rm rad}$ of the dipole transition matrix
element $\mu=\mu_{\rm rad}\times\mu_{\rm sph}$ with
\begin{equation}
 \mu_{\rm rad} =
 \left\langle \frac{1}{r}U_{5,P,3/2}
 \right| er \left|
 \frac{1}{r}U_{\rydn,\ell,j} \right\rangle
 \,.
\end{equation}
For large principal quantum numbers it scales as $\mu_{\rm rad}
\simeq C_\ell \, {(\rydn^*)}^{-3/2}$, where $\rydn^*$ denotes the
principal quantum number reduced by the appropriate quantum
defect. For \rydn$\geq$30 we find $C_0$=4.508\,a.u. and
$C_2$=8.475\,a.u. while the spherical matrix elements for the
stretched transitions as depicted in Fig.~\ref{fig:levelscheme}
are $\mu_{sph}=\sqrt{1/3}$ for $\ell=0$(S) and $\sqrt{2/5}$ for
$\ell=2$(D). Table~\ref{tbl:TheoVsExp} shows the comparison
between the experimental values and the theoretical prediction.
While there is excellent agreement for the 31S state, the measured
value for the 30D state is slightly smaller than expected. The
deviation is most likely caused by electric stray fields which
induce an admixture of other $m_J$-states effectively reducing the
spherical matrix element. This effect does not perturb the
S$_{1/2}$ state as $m_J=1/2$ is the only dipole coupled state and
the Stark effect on the intermediate 5P$_{3/2}$ state is
negligible.

\subsection{Redistribution to other states}

\begin{figure}
 \begin{indented}
  \item[]\includegraphics[width=.65\textwidth]{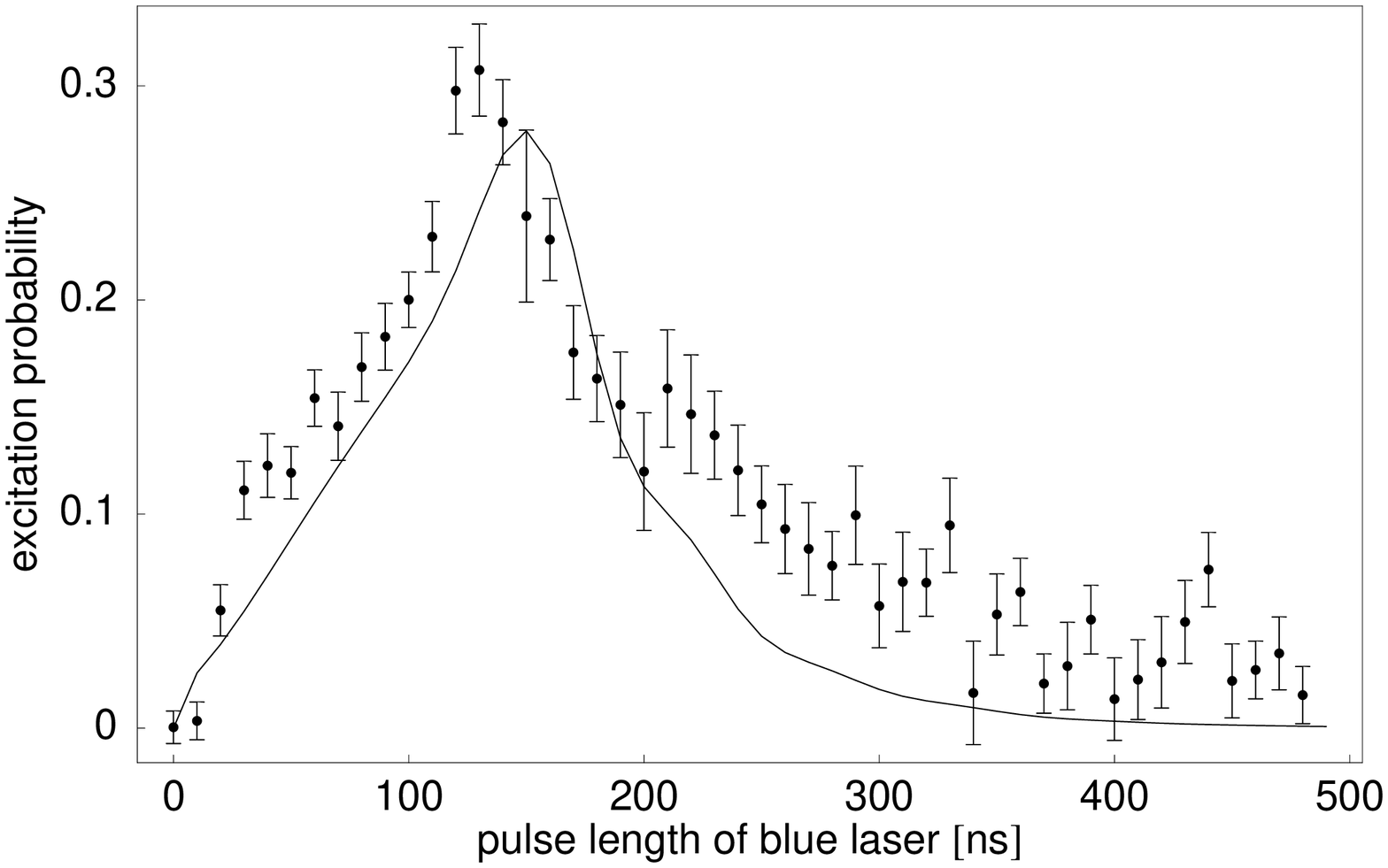}
 \end{indented}
\caption{Quenching of Rydberg atoms by stimulated emission. At
150\,ns, the laser at 780\,nm for the lower transition is turned
off, and Rydberg atoms in the 32S$_{1/2}$ are stimulated to the
5P$_{3/2}$ state by the laser light at 480\,nm. Experimental
values of the total Rydberg excitation with the corresponding
statistical error are shown as dots. The solid line shows the
excitation probability of the 32S$_{1/2}$ state as predicted by a
model calculation with no free parameters. The scaling factor for
the MCP signal was determined by comparing the corresponding Rabi
oscillation to a simulated curve as in Fig. \ref{fig:expRabi}.}
 \label{fig:quenching}
\end{figure}

Redistribution of Rydberg atoms to other states due to population
transfer processes has already been observed to reduce the
efficiency of de-excitation (quenching) in a macroscopic cloud of
Rydberg atoms~\cite{cubel05}. By measuring the quenching for the
32S$_{1/2}$ state as shown in Fig.~\ref{fig:quenching} this
process can be ruled out for the small mesoscopic ensemble
described here. After 150\,ns of excitation the 780\,nm laser is
turned off, while the laser for the upper transition is left on.
Due to the short lifetime of the intermediate 5P$_{3/2}$ state,
population inversion is created. The blue laser stimulates
transitions of the Rydberg state to the intermediate state, which
then quickly decays to the ground state. Redistribution processes
would lead to population in other Rydberg states which are not
coupled by the blue laser light. As these atoms are still field
ionised, they should result in a finite MCP signal even for long
quenching times. As Fig.~\ref{fig:quenching} shows, the total
number of Rydberg atoms is almost fully quenched by stimulated
emission. In addition, the temporal evolution of the Rydberg
signal is in good agreement with the solution of the Bloch
equations for a pure three-level system shown as a solid line.
Redistribution of Rydberg population can therefore be ruled out as
a major cause of decoherence in the mesoscopic ensemble.

\subsection{Sources for decoherence}
\label{sec:dephasing}

Besides redistribution of Rydberg atoms to other states, three
factors cause the reduced contrast and the damping of Rabi
oscillations:  the residual inhomogeneity of the flat-top
intensity profile, the small but finite admixture of the
short-lived intermediate 5P$_{3/2}$ state, and the finite
linewidth of the laser source. Note that only the last two effects
are sources of decoherence for all atoms while the first effect
only represents ensemble averaging with full coherence for each
individual atom.

The influence of the intermediate state and the residual intensity
distribution on the damping of the Rabi oscillations are simulated
by solving the optical Bloch equations (OBE) for the 3-level
system and averaging over different blue Rabi frequencies
according to the measured intensity distribution shown in
Fig.~\ref{fig:levelscheme}. The only free parameters are the
detection efficiency of the MCP and the average Rabi frequency of
the upper transition, which agrees well with theory (see
Section~\ref{sec:abinitio}). The resulting simulations are
depicted as dotted and dashed lines in Fig.~\ref{fig:expRabi}
showing damped Rabi oscillations with reduced contrast. The dotted
line is a simulation for a single Rabi frequency and shows a
reduced contrast due to the admixture of the intermediate state
which can be decreased by increasing the detuning $\delta$ which
in turn results in a smaller effective Rabi frequency. The dashed
line additionally incorporates an averaging over different Rabi
frequencies corresponding to the measured residual intensity
fluctuation in the flattop beam profile. The inhomogeneous
distribution of Rabi frequencies alone does not fully explain the
reduced measured contrast.

Indeed the main contribution to the reduced contrast is the finite
bandwidth of the laser sources. As the admixture of the
intermediate level is a small we can incorporate the laser
bandwidth $\gamma$ into the simulation by solving the optical
Bloch equation for a two-level system with a linewidth of $\gamma$
and an effective Rabi frequency of $\Omega_{\mathrm{eff}} =
\omega\,\Omega\,/\,2\delta$ where $\omega$ ($\Omega$) denotes the
Rabi frequency of the lower (upper) transition and $\delta$
denotes the detuning from the intermediate level. This is a good
approximation for $\delta\gg\omega,\,\Omega$ \cite{shore90}. We
have again averaged over the different values of
$\Omega_{\mathrm{eff}}$ as present in the imperfect flat top. The
corresponding model calculation for an effective incoherent
excitation bandwidth of $\gamma=2\pi\times2.4\,$MHz is shown as a
solid line in Fig.~\ref{fig:expRabi} in excellent agreement with
the experimental data. The experimental value for $\gamma/(2\pi)$
agrees well with the specified bandwidths of 2\,MHz for the blue
and 1\,MHz for the red excitation laser. In addition we have
performed beating measurements with two comparable, independent
lasers at 780\,nm which yielded a combined bandwidth of 2\,MHz, in
good agreement with the above values.

\section{Van der Waals blockade}

\begin{figure}
 \begin{indented}
  \item[]\includegraphics[width=.65\textwidth]{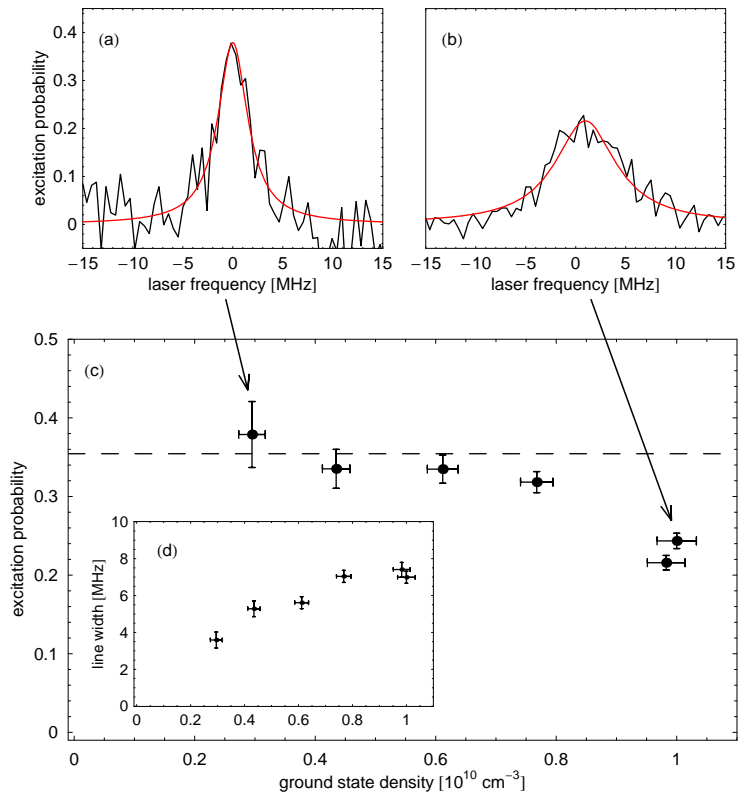}
 \end{indented}
\caption{Excitation blockade for a mesoscopic ensemble of atoms in
the 75S state. The upper graphs show excitation spectra for the
highest (a) and lowest (b) investigated densities together with
Lorentzian fits. The higher density is clearly accompanied by
smaller excitation probability and a significant line broadening.
The excitation probabilities are calibrated by measurements at low
\rydn, which results in single atom excitation probabilities. (a)
shows the maximum excitation for different ground state densities.
The dashed line corresponds to the theoretical excitation
probability for non-interacting atoms. For higher densities the
Rydberg-Rydberg interaction energy rises and moves the excitation
out of resonance reducing the excitation probability on resonance.
The corresponding line broadening (shown in (d) with an identical
x-axis) also reflects the increasing interaction energy.}
 \label{fig:Blockade}
\end{figure}

The excitation presented so far has been performed in a regime
with negligible interactions at small~\rydn. For large values of
$\rydn\gtrsim60$ we do not observe Rabi oscillations as the Rabi
frequency, scaling with $\rydn^{-3/2}$ becomes smaller than the
dephasing rate. On the other hand these higher lying states
exhibit stronger Rydberg-Rydberg interactions which can result for
example in a van der Waals blockade. Fig.~\ref{fig:Blockade}(c)
shows how the excitation probability of the 75S state depends on
the ground state density. For a single (or non-interacting) atom
we can solve the optical Bloch equation for the independently
determined experimental parameters and obtain a value represented
by the dashed line. For low densities the measured excitation
probability agrees well with the prediction. With increasing
density we observe an increasing suppression of the excitation
probability. It is accompanied by a significant line broadening
which is shown as inset in Fig.~\ref{fig:Blockade}(d) as well as
in the exemplary spectra in Fig.~\ref{fig:Blockade}(a) and (b).
For small densities the measured linewidth agrees well with a
convolution of our laser bandwidth (2.4\,MHz) and saturation
broadening (1.7\,MHz).

The suppression of excitation marks the onset of the dipole
blockade. The underlying interaction has van der Waals character
with a $C_6/R^6$ dependence on the interatomic spacing $R$ in
contrast to dipole-dipole interaction with $R^{-3}$ character,
which was originally proposed~\cite{lukin01}. In comparison this
means a shorter interaction range and in fact we see the
suppression due to vdW interaction only for high principal quantum
numbers \rydn~as the vdW coefficient scales as $C_6 \propto
\rydn^{11}$.

These results complement our earlier measurements with macroscopic
atom clouds~\cite{singer04} which showed a much stronger
broadening at comparable densities and excitation suppression. In
fact the observed broadening in the macroscopic cloud cannot be
explained by mere Rydberg-Rydberg interactions~\cite{ates07PRA}.
One possible explanation may be ions that are produced by
interaction-induced collisions between the cold atoms after
acceleration by the strong van der Waals forces between Rydberg
atoms~\cite{amthor07,amthor07b}. For the mesoscopic system
investigated here we have ruled out the presence of ions directly
by selective field ionisation. This is also expected for the small
number of involved atoms and the short excitation times.

\section{Conclusion}

We have observed Rabi oscillations of a mesoscopic cloud of about
100 atoms in the coherent excitation of Rydberg states out of an
ultracold atom cloud. The reduced contrast of the Rabi oscillation
is traced back to residual intensity variations and the finite
bandwidth of our laser system. While the latter can simply be
solved improving the frequency stability of the laser systems, one
option to bypass the first limitation is the use of liquid crystal
elements~\cite{bergamini04} to dynamically optimise the
diffractive pattern for better results.

In addition we have demonstrated the van der Waals blockade at
large principal quantum numbers for a mesoscopic cloud. This
indicates the onset of a local blockade as has been observed
before for macroscopic
clouds~\cite{tong04,singer04,cubel05a,vogt06,vogt07,heidemann07}.
The full blockade regime can be identified by an increased Rabi
frequency $\sqrt{N}\,\Omega$, where $N$ and $\Omega$ denote the
number of atoms and the atomic Rabi frequency,
respectively~\cite{lukin01}. The experiments presented here
constitute a necessary prerequisite for the direct observation of
this coherent many-body phenomenon. In future work we will use a
number of options to reach this regime by increasing the
interaction strength through higher densities~\cite{heidemann07}
and using either resonant~\cite{vogt06} or permanent~\cite{vogt07}
dipole-dipole interactions. Smaller excitation
volumes~\cite{johnson07} will provide ensembles where even the
interaction between the farthest separated atoms dominates all
other energy scales.

\ack

We thank C.~Pruss and W.~Osten from the Institute for Technical
Optics at the University of Stuttgart for the design and
production of the beam shaping element. The project is supported
by the Landes\-stiftung Baden-W\"urttemberg within the ''Quantum
Information Processing'' programme, and by a grant from the
Ministry of Science, Research and Arts of Baden-W\"urttemberg (Az:
24-7532.23-11-11/1).

\section*{References}

\end{document}